\documentclass[prb,reprint,superscriptaddress,
amsmath,amssymb,aps,]{revtex4-2}
\usepackage{array}
\usepackage{amsmath}
\usepackage{graphicx}
\usepackage{dcolumn}
\usepackage{bm}
\usepackage{float}
\usepackage{hyperref}
\usepackage{physics}
\usepackage{subcaption}
\usepackage{booktabs}
\hypersetup{colorlinks,linkcolor={blue},citecolor={blue},urlcolor={blue}}  
\usepackage{xtab,afterpage}
\usepackage[utf8]{inputenc}
\DeclareUnicodeCharacter{03C3}{\ensuremath{\sigma}}
\makeatletter
\def\LT@LR@e{\LTleft\z@   \LTright\z@}%
\makeatother

\begin{document}

\preprint{APS/123-QED}

\title{Classical Benchmarks of a Symmetry-Adapted Variational Quantum Eigensolver for Real-Time Green’s Functions in Dynamical Mean-Field Theory}


\author{Aadi Singh}
\affiliation{Center for Computation and Technology, Louisiana State University, Baton Rouge, LA 70803, USA}
\affiliation{Department of Physics and Astronomy, Louisiana State University, Baton Rouge, LA 70803, USA}

\author{Chakradhar Rangi}
\affiliation{Department of Physics and Astronomy, Louisiana State University, Baton Rouge, LA 70803, USA}

\author{Ka-Ming Tam}
\affiliation{Center for Computation and Technology, Louisiana State University, Baton Rouge, LA 70803, USA}
\affiliation{Department of Physics and Astronomy, Louisiana State University, Baton Rouge, LA 70803, USA}

\date{\today}

\begin{abstract}
We present a variational quantum eigensolver (VQE) approach for solving the Anderson Impurity Model (AIM) arising in Dynamical Mean-Field Theory (DMFT). Recognizing that the minimal two-site approximation often fails to resolve essential spectral features, we investigate the efficacy of VQE for larger bath discretizations while adhering to near-term hardware constraints. We employ a symmetry-adapted ansatz enforcing conservation of particle number $(N)$, spin projection $(S_z=0)$, and total spin $(S^2=0)$ symmetry, benchmarking the performance against exact diagonalization across different interaction strengths using bath parameters extracted from the DMFT self-consistency loop. For a four-site model, the relative error in the ground state energy remains well below $0.01\%$ with a compact parameter set $(N_p \le 30)$. Crucially, we demonstrate that the single-particle Green's function—the central quantity for DMFT—can be accurately extracted from VQE-prepared ground states via real-time evolution in the intermediate to strong interaction regimes. However, in the weak interaction regime, the Green's function exhibits noticeable deviations from the exact benchmark, particularly in resolving low-energy spectral features, despite the ground state energy showing excellent agreement. These findings demonstrate that VQE combined with real-time evolution can effectively extend quantum-classical hybrid DMFT beyond the two-site approximation, particularly for describing insulating phases. While this approach offers a viable pathway for simulating strongly correlated materials on near-term devices, the observation that accurate ground state energy does not guarantee accurate dynamical properties highlights a key challenge for applying such approaches to correlated metals.
\end{abstract}\maketitle

\section{Introduction}

Unlocking the properties of strongly correlated electron systems—which underpin the physics of high-temperature superconductors, Mott insulators, and advanced semiconductors—remains a persistent challenge after decades of effort \cite{Imada_RMP_1998, Lee_RMP_2006}. These complex systems are typically described by lattice Hamiltonians in which electrons hop between sites while experiencing strong on-site Coulomb repulsion. Solving such many-body models on a macroscopic lattice is computationally intractable due to the exponential scaling of the Hilbert space with system size. While Density Functional Theory (DFT) has been tremendously successful for weakly interacting systems \cite{Kohn_Sham_1965,Hohenberg_Kohn_1964,Perdew_Burke_Ernzerhof_1996}, its application to strongly correlated materials is limited, particularly for systems with narrow $d$- or $f$-bands where Coulomb repulsion competes strongly with kinetic energy \cite{DFT+DMFT_RMP}.

To address this, various methods have been proposed to improve upon standard DFT by incorporating strong correlation effects \cite{DFT+DMFT_RMP}. One particularly successful strategy is Dynamical Mean-Field Theory (DMFT)\cite{Georges_Kotliar_1992,DMFT_RMP,Muller_Hartmann_1989a,Muller_Hartmann_1989b,Metzner_Vollhardt_1989,Jarrell_1992}, which maps the intractable lattice problem onto a self-consistent quantum impurity model. Here, the complex lattice Hamiltonian is replaced by a single interacting site coupled to a non-interacting bath. This construction captures the essential local temporal fluctuations by treating the on-site interactions exactly, while approximating spatial correlations as a mean field.

Despite this reduction to a local problem, the framework successfully captures key characteristics of strongly correlated systems, most notably the metal-insulator phase transition driven by increasing electron interaction strength \cite{DMFT_RMP}. Formally, DMFT becomes exact in the limit of infinite dimensions. In practice, however, the choice of impurity solver determines the achievable accuracy and system size. Widely used action-based solvers such as continuous-time quantum Monte Carlo (CT-QMC) handle the non-interacting bath directly without discretization, but they frequently suffer from the fermion sign problem in multi-orbital or non-equilibrium settings \cite{CTQMC_RMP, Werner_PRB_2009}. Conversely, Hamiltonian-based solvers such as exact diagonalization (ED) require discretizing the bath into a finite set of orbitals, at which point the exponential growth of the Hilbert space re-emerges with increasing bath size \cite{Caffarel_Krauth_1994, Capone_etal_2007}. 

Building on these solver paradigms, stochastic sampling methods have evolved from the early Hirsch-Fye algorithm, which uses a Trotter decomposition, to modern continuous-time methods based on sampling diagrammatic series expansions \cite{Hirsch_Fye_1986}. Hamiltonian-based methods range from direct exact diagonalization (e.g., Lanczos and Davidson algorithms) and the Numerical Renormalization Group (NRG) to the modern Density Matrix Renormalization Group (DMRG) and other tensor network approaches \cite{Anisimov_Zaanen_Andersen_1991, Vollhardt_2011,Bulla_Costi_Pruschke_2008,Peters_etal_2006,Ganahl_etal_2015,Wolf_etal_2015,Bauernfeind_etal_2017,Weh_etal_2021}. 

Beyond these traditional frameworks, recent exploratory methods based on machine learning have been proposed \cite{Arsenault_etal_2014,Rigo_etal_2020,Sheridan_etal_2021,Walker_etal_2022,Sturm_etal_2021}, alongside data-compression techniques such as tensor train methods \cite{Erpenbeck_etal_2023}. Furthermore, generalizations of the DMFT method that include spatial correlations often involve multiple impurity systems, for which generic, efficient numerical solvers are not yet available \cite{DCA_RMP,DCA_Fotso2012}. The resulting exponential growth in computational complexity creates massive data storage and processing requirements, posing a difficult task for large systems even with high-end modern hardware.

A promising direction for overcoming these computational barriers has emerged with the advent of quantum computing. Unlike classical systems, quantum computers naturally encode correlated fermionic wavefunctions, theoretically circumventing the exponential memory scaling through the principle of superposition and entanglement. This makes them an attractive platform for tackling quantum impurity models, enabling simulations of larger and more complex systems than feasible today. In the Noisy Intermediate-Scale Quantum (NISQ) era, the Variational Quantum Eigensolver (VQE) has emerged as the go-to method for ground-state preparation \cite{Peruzzo}. Early studies have already demonstrated the feasibility of using VQE ground state to extract Green’s functions for the Anderson Impurity Model with a single bath site, see e.g. Ref. \cite{Ayral_2025} for the review. These proof-of-concept successes suggest a feasible pathway for scaling up to larger clusters, which offer significantly improved spectral resolution for resolving features like Hubbard bands. If VQE can effectively resolve the Green’s function across weakly to strongly correlated regimes for parameters appropriate for DMFT, it could establish a robust framework for quantum-classical hybrid DMFT \cite{Ayral_2025,Baul_etal_2025,Kreula_etal_2016,Besserve_Ayral_2022,Kreula_Clark_Jaksch_2016,Rungger_etal_2019,Jaderberg_etal_2020,Yao_etal_2021,Steckmann_etal_2023,Jamet_etal_2021,Jamet_Agarwal_Rungger_2022,Ehrlich_Urban_Elsasser_2024,Jamet_etal_2025,Baul_etal_2025,Selisko_etal_2024,Jones_etal_2025,Dhawan_Zigid_Motta_2023,Perez_etal_2025,hogan2025,Karabin_etal_2026} that can eventually be extended to multi-orbital systems and realistic material simulations \cite{DFT+DMFT_RMP}.

In this work, we systematically benchmark a symmetry adapted VQE approach for the Anderson Impurity Model beyond the two-site limit. We employ an ansatz that explicitly enforces particle number, spin along $z$-axis, and total spin conservation. To access the dynamical properties required for DMFT, we compute the single particle Green's function via real time evolution using the Suzuki-Trotter decomposition \cite{Trotter_1959,Suzuki_1976,Suzuki_1977}. We evaluate the method on a four-site cluster with realistic DMFT parameters. While VQE yields ground state energies with $<0.01\%$ error, we find that the dynamical Green's function exhibits significant deviations in the weak interaction regime. These findings highlight that variational accuracy in ground-state energy does not guarantee the wavefunction fidelity necessary for reliable real-time dynamics.

The paper is organized as follows. In Section II, we describe the model Hamiltonian and the theoretical background. We also detail the construction of the symmetry-preserving VQE ansatz and review the Suzuki-Trotter method for calculating the retarded Green's function. In Section III, we present the convergence results for the VQE ground state energy and utilize the optimized wavefunction to compute the real-time Green's function. Finally, we conclude in Section IV and provide perspectives on future studies.

\section{Anderson Impurity Model and VQE wavefunction}

\subsection{Anderson Impurity Model and DMFT}

The Anderson Impurity model (AIM) and the related Kondo model are cornerstones of condensed matter physics, serving as the simplest nontrivial examples of interacting many-body systems \cite{doi:10.1098/rspa.1963.0204}. Crucially, the AIM forms the core of DMFT, which has become a powerful tool for studying strongly correlated materials. DMFT maps the lattice problem onto a self-consistent single-impurity model in the limit of infinite lattice coordination number (or infinite dimensions). In this limit, spatial fluctuations are suppressed and the self-energy-which captures all many-body correlation effects beyond the mean-field level-becomes purely local, losing its momentum dependence. The lattice is effectively replaced by one interacting impurity site coupled to a non-interacting fermionic bath, whose hybridization function encodes the mean-field effect of all other sites \cite{Muller_Hartmann_1989a,Muller_Hartmann_1989b,Metzner_Vollhardt_1989}. This hybridization function is not known a priori; instead, it must be determined self-consistently through an iterative procedure. The self-consistency condition requires solving the impurity problem for a given hybridization function to obtain the impurity Green's function, from which a new hybridization function is constructed via the Dyson equation and the lattice dispersion relation. This self-consistency loop continues until convergence is achieved, at which point the impurity and lattice descriptions are mutually compatible, and the DMFT equations are satisfied. 

The central computational challenge in DMFT is therefore solving the effective AIM accurately and efficiently. In this work, we follow the standard exact-diagonalization (ED) approach by discretizing the continuous bath into a finite number of bath sites \cite{Liebsch_Ishida_2011}. The resulting finite-size impurity model consists of one interacting impurity site coupled to $N_b$ non-interacting bath sites, with hopping amplitudes $t_{i,j}$ between the impurity and bath sites (and between bath sites in chain geometry) and local on-site energies $\epsilon_i$ on the bath sites. We adopt the chain geometry for the bath, where the impurity is attached to one end of a linear chain of bath sites connected by nearest-neighbor hoppings and local energies. This chain representation is particularly advantageous when fermionic operators are mapped to spin-1/2 operators via the Jordan-Wigner transformation, as the nearest-neighbor structure minimizes the number of long-range strings in the Pauli operators, simplifying both classical exact diagonalization and quantum circuit implementation. The chain geometry can be obtained exactly from the original star geometry (where the impurity couples directly to all bath sites) via the Lanczos algorithm, which tridiagonalizes the bath Hamiltonian. In this work, we therefore focus on the linear chain representation throughout our calculations.

The Hamiltonian for the discretized effective AIM in the linear chain geometry is given by:
\begin{gather}
    H_{AIM} = \sum_{\sigma=\uparrow,\downarrow} \Biggl(-\sum_{i=0}^{N_b-1} t_{i,i+1} \left( c^{\dagger}_{i,\sigma} c_{i+1,\sigma} + c^{\dagger}_{i+1,\sigma} c_{i,\sigma} \right) + \nonumber \\ \sum_{i=1}^{N_b} \epsilon_i n_{i,
    \sigma} - \mu n_{0,\sigma}\Biggr) + Un_{0\uparrow} n_{0\downarrow},
    \label{eq:AIM}
\end{gather}
where $c_{i,\sigma}^\dagger$ ($  c_{i,\sigma}  $) are fermionic creation (annihilation) operators satisfying the standard fermionic anti-commutation relations, $n_{i,\sigma}=c_{i,\sigma}^\dagger c_{i,\sigma}$ is the number operator at site $i$ with spin $\sigma $, $t_{i,i+1}$ are the nearest-neighbor hopping amplitudes along the chain, $\epsilon_i$ are the on-site energies of the bath sites, $\mu$ is the chemical potential, $U$ is the local Hubbard repulsion on the impurity (site 0), and $N_b$ is the number of bath sites. We shall represent the total number of sites in the linear chain by $N=N_b+1$. 

To simulate the fermionic Hamiltonian $H_{AIM}$	on a digital quantum computer, we map the fermionic operators to qubit operators. We employ the standard Jordan-Wigner (JW) transformation \cite{JW_1928}, which encodes the occupation state of each spin-orbital into the state of a qubit ($\ket{0} \equiv$ empty, $\ket{1} \equiv$ occupied). We utilize a blocked mapping scheme where the first $N$ qubits encode the spin-up orbitals (indices $0$ to $N-1$) and the subsequent $N$ qubits encode the spin-down orbitals (indices $N$ to $2N-1$). A crucial advantage of the linear chain geometry chosen in Eq. (\ref{eq:AIM}) is that the nearest-neighbor fermionic hopping maps directly to local 2-qubit interactions under the JW transformation. This avoids the non-local Pauli strings that arise in star geometries, significantly reducing the gate count and circuit depth required for simulation.

Applying this transformation to Eq. (\ref{eq:AIM}) yields the following qubit Hamiltonian acting on $2N$ qubits:
\begin{eqnarray}
H_{\text{qubit}} =  -\frac{1}{2} \sum_{i=0}^{N-1} t_{i,i+1} \Big( X_i X_{i+1} + Y_i Y_{i+1} \nonumber \\  + X_{N+i} X_{N+i+1} + Y_{N+i} Y_{N+i+1} \Big) \nonumber \\
 + \frac{U}{4} Z_0 Z_N - \sum_{i=1}^{N} \frac{\epsilon_i}{2} (Z_i + Z_{N+i}) \nonumber \\
 - \frac{1}{2} \left( \mu + \frac{U}{2} \right) (Z_0 + Z_N),
\label{eq:AIM_Sigma}
\end{eqnarray}
where $X_i, Y_i, Z_i$ denote the Pauli spin operators for the $x, y, z$ directions on the i-th qubit. The first set of hopping terms acts on the spin-up block, and the second set acts on the spin-down block, with the Coulomb interaction $U$ coupling the impurity qubits at indices $0$ and $N$. The bath parameters $\{t_{i,i+1},\epsilon_i\}$ are determined by solving the self-consistent DMFT equation. 

\subsection{Symmetry-Preserving VQE Ansatz}\label{section:IIB}
We employ the VQE algorithm to construct and optimize the ground state wavefunction for the Anderson impurity model [Eq (\ref{eq:AIM_Sigma})]. This hybrid algorithm leverages the strengths of both quantum and classical processors. First, we prepare a parameterized quantum circuit representing the trial wavefunction and measure its energy expectation value on quantum hardware. A classical optimizer then processes these measurements to update the parameters, minimizing the energy to locate the ground state. This feedback loop allows the VQE to prepare wavefunctions on the quantum processor while offloading the minimization task to a classical computer.

Constructing the wavefunction with the correct symmetry is crucial for achieving high-fidelity ground states, as suggested in recent studies \cite{Jones_etal_2025}. The Hamiltonian in Eq. (\ref{eq:AIM}) commutes with the total particle number operator $\hat{N}$ and the total spin projection $\hat{S}_z$, making these conserved quantities. Furthermore, for the Anderson impurity model at half-filling with repulsive interactions, Lieb's theorem \cite{Lieb_1989} guarantees that the ground state is a unique spin singlet ($S=0$). Although originally formulated for the Hubbard model on a bipartite lattice, the theorem applies here because the linear chain geometry is bipartite and the interaction terms are semi-positive definite. Consequently, we restrict our variational search to the subspace defined by total particle number conservation ($N$), total spin projection ($S_z=0$), and total spin ($\hat{S}^2$) symmetry.

\begin{figure*}[t]
\centering\includegraphics[width=\textwidth]{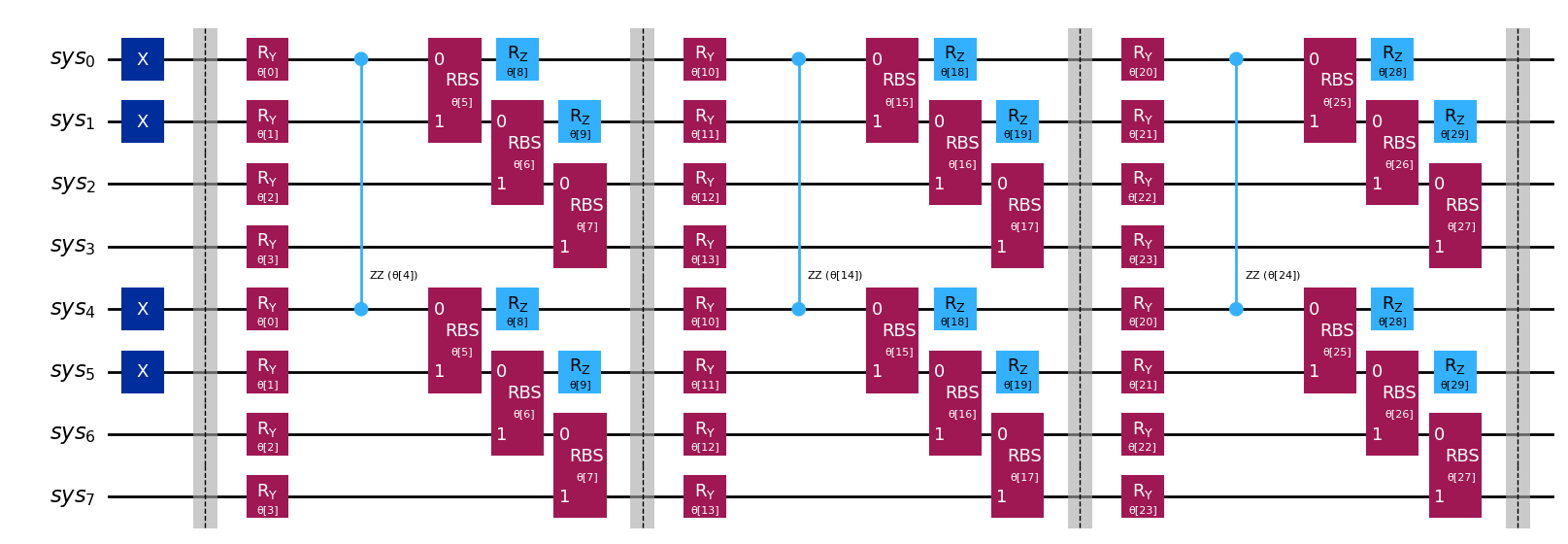} 
    \caption{ Quantum Circuit for the variational wavefunction. $R_y$ and $R_z$ denote rotational gates along the $Y$ and $Z$ axes, respectively. $\theta[i]$ are the rotation angles, which are treated as parameters to be optimized. This schematic illustrates the specific implementation for $N=4$ sites with 3 layers; with each layer containing 10 parameters, resulting in a total of  tunable parameters for the entire circuit.}
    \label{fig:VQE_circuit}
\end{figure*}



\begin{figure}[ht]
    \centering    
    \includegraphics[width=0.33\textwidth]{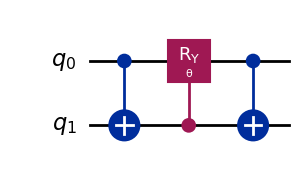}
    \caption{The gate connecting two qubits to implement the hopping term, also denoted as the Reconfigurable Beam Splitter (RBS) gate.}
    \label{fig:RBS}
\end{figure}

We implement these constraints using the symmetry-preserving ansatz. While this construction is generalizable to any number of bath sites, we illustrate the specific circuit implementation for $N=4$ sites (requiring $2N=8$ qubits) in Fig. \ref{fig:VQE_circuit}. The circuit begins by creating an initial reference state: a product state of two doubly occupied sites (the impurity and the first bath site). This state has an equal number of spin-up and spin-down electrons and is inherently a singlet. To transform this reference state into the trial ground state while preserving its symmetry, we employ specific parameterized gates:

\begin{enumerate}
    \item We begin by applying a Pauli-Y rotation, $R_y(\theta) = e^{-i\theta Y/2}$ to every site. To strictly maintain the spin rotational symmetry ($S^2=0$), we enforce parameter sharing: the qubits representing spin-up electrons and their corresponding spin-down counterparts share identical rotation parameters.
    \item Next, we apply a coupling gate to the impurity site to account for the on-site Coulomb repulsion. Since the interaction term $U\hat{n}_{0\uparrow}\hat{n}_{0\downarrow}$ maps to a $ZZ$ rotation under the JW transformation, we implement this as $\exp(-i\theta Z_0 Z_N)$. This diagonal operation naturally commutes with all symmetry operators.
    \item We then apply Reconfigurable Beam Splitter (RBS) gates to connect the sites linked by the hopping terms. As shown in Fig. \ref{fig:RBS}, this gate is composed of two CNOT gates and one controlled-$R_y$ rotation (which can be further decomposed into two CNOTs and single-qubit $R_y$ rotations). We again enforce parameter sharing between spin-up and spin-down hopping gates to preserve spin symmetry.
    \item Finally, we add a single $R_z$ rotation gate specifically on the impurity site and the first bath site connected to it.
\end{enumerate}
This complete sequence constitutes one variational layer. For the specific $N=4$ model illustrated here, this architecture results in a total of 10 variational parameters per layer.

\subsection{Calculating Green’s function}

The single-particle Green's function is the central quantity for describing many-body systems, providing direct access to the spectral function and density of states. Once the ground state energy is minimized via VQE, the next critical step is to determine the system's excitation spectrum. These excitations are described by adding or removing a particle from the ground state at different times, a process mathematically captured by the Green's function. In this work, we focus on the retarded Green's function, which encodes the causal dynamical response of the system.

We calculate the retarded Green’s function in the time domain via Trotter propagation and the Hadamard test, following protocols established in Refs. \cite{Kreula_etal_2016,Keen_etal_2020}. A known trade-off of this approach is that the circuit depth scales linearly with the simulation time, necessitating deep circuits for long-time evolution. However, the critical advantage is that the algorithmic error is controlled by the Trotter time step size, which allows for a systematic improvement. Alternative quantum algorithms for Green's functions include subspace expansion methods, such as Krylov-space approaches \cite{Greene_etal_2024,Jamet_Agarwal_Rungger_2022} or physically inspired excitation coupled-cluster methods \cite{Keen_etal_2021}. While these methods avoid deep circuits by capturing the dominant spectral weight within a small subspace, their error is not strictly controlled by a small parameter, making accuracy dependent on the choice of subspace. Other proposed techniques include orthogonal polynomial expansions for the density of states \cite{Summer_etal_2024} and variational time propagation schemes \cite{Gomes_Young_deJong_2023,Kanasugi_etal_2023,Endo_etal_2020,Claisse_etal_2025}.

Since our primary goal is to benchmark the quality of the VQE wavefunction for DMFT applications, we prioritize the controlled error limits of real-time Trotterization over the heuristic advantages of subspace methods. We assume the system is in a paramagnetic state where spin rotational symmetry is preserved. Consequently, we only need to measure the Green's function for a single spin component. The retarded Green's function ($G^{R}$(t)) is defined in terms of the greater ($G^{>}$) and lesser ($G^{<}$) Green's functions as \cite{Freericks_2019}:

\begin{align}
     G^{R}_{imp}(t) &=\theta(t)(G^{>}_{imp}(t)-G^{<}_{imp}(t))  \nonumber \\ 
     &=-\mathrm{i} \theta(t)(\langle c_{0 \sigma}(t)c^{\dagger}_{0\sigma}(0)\rangle+\langle c^{\dagger}_{0 \sigma}(0)c_{0\sigma}(t)\rangle).
   \label{eq:GR}
 \end{align} 
After the JW transformation, the greater and lesser components can be expanded into measurable Pauli expectation values, give as: \cite{Kreula_etal_2016,Endo_etal_2020}:
\begin{eqnarray}
    G^{>}_{imp}(t)&=&-\frac{\mathrm{i}}{4} [\langle \hat{U}^{\dagger}(t)X_{0}\hat{U}(t)X_{0} \rangle 
    +\mathrm{i} \langle \hat{U}^{\dagger}(t)X_{0}\hat{U}(t)Y_{0} \rangle \nonumber\\
    &-&\mathrm{i}\langle \hat{U}^{\dagger}(t)Y_{0}\hat{U}(t)X_{0} \rangle 
    + \langle \hat{U}^{\dagger}(t)Y_{0}\hat{U}(t)Y_{0}\rangle],
    \label{eq:G_larger}
\end{eqnarray}
and 
\begin{eqnarray}
     G^{<}_{imp}(t)&=&\frac{\mathrm{i}}{4} [\langle X_{0}\hat{U}^{\dagger}(t)X_{0}\hat{U}(t) \rangle \nonumber 
     -\mathrm{i} \langle X_{0}\hat{U}^{\dagger}(t)Y_{0}\hat{U}(t) \rangle \nonumber \\
     &+&\mathrm{i} \langle Y_{0}\hat{U}^{\dagger}(t)X_{0}\hat{U}(t) \rangle  
     + \langle Y_{0}\hat{U}^{\dagger}(t)Y_{0}\hat{U}(t) \rangle ].
         \label{eq:G_smaller}
\end{eqnarray}

In quantum computing implementations, those expectation values can be measured by introducing an ancilla bit to perform the Hadamard test \cite{Kreula_etal_2016}. 



\section{Results}

We evaluate the quality of the VQE wavefunction by examining both the ground state energy and the real-time retarded Green's function. The calculation of the Green's function implicitly probes all excited states of the system, providing a stringent test of the VQE ansatz beyond simple ground state properties.

\subsection{Model Setup and Parameters}

\begin{figure}[b]
    \centering
    \includegraphics[width=9cm]{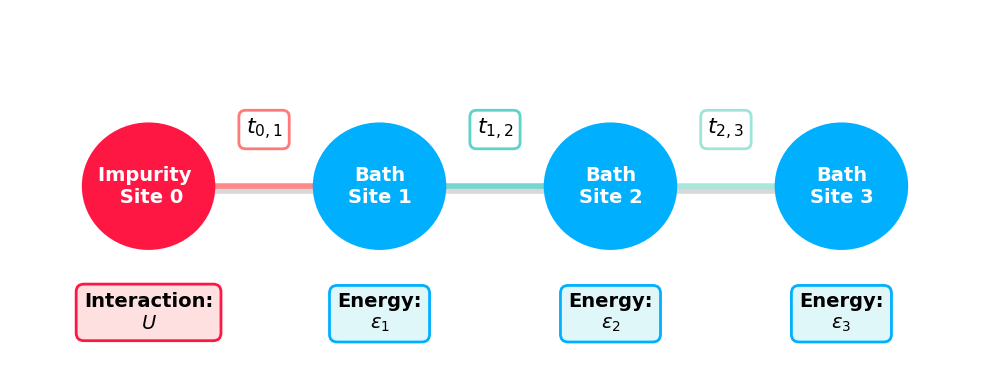}
    \caption{Schematic of the 4-site one-dimensional chain for the Anderson impurity model, consisting of a single impurity site (site 0) coupled to three bath sites.}
    \label{fig:aim_4sites}
\end{figure}

\begin{table}[ht]
    \centering
    \begin{tabular}{c c c c}
        \toprule
        $U$ & $t_{0,1}$ & $t_{1,2}$ & $\epsilon_1$ \\
        \midrule
        4 & 0.2164 & 0.3646 & 0.1863 \\
        6 & 0.1549 & 0.5177 & 0.4359 \\
        8 & 0.1405 & 0.6850 & 0.6390 \\
        \bottomrule
    \end{tabular}
    \caption{Model parameters for the 4-site Anderson Impurity Model at varying interaction strengths $U$, obtained from converged DMFT calculations. All parameters satisfy the particle-hole symmetry constraints: $t_{1,2} = t_{2,3}$, $\epsilon_2 = 0$, and $\epsilon_3 = -\epsilon_1$.}
    \label{table:4site_parameters}
\end{table}

We employ the symmetry-adapted VQE circuit, depicted in Fig. \ref{fig:VQE_circuit} and described in Section \ref{section:IIB}, to solve the Anderson impurity model for a 4-site cluster consisting of 1 impurity site and 3 bath sites, as illustrated in Fig. \ref{fig:aim_4sites}. The model parameters—hopping amplitudes $t_{0,1}, t_{1,2}, t_{2,3}$ and on-site energies $\epsilon_1, \epsilon_2, \epsilon_3$—are obtained from converged DMFT self-consistency calculations for the single-band Hubbard model at half-filling on the Bethe lattice. We investigate three interaction strengths representing different correlation regimes: $U = 4$ (correlated metal), $U = 6$ (around metal to insulator transition), and $U = 8$ (Mott insulator). The corresponding parameters are listed in Table \ref{table:4site_parameters}. Due to particle-hole symmetry at half-filling, the parameters satisfy the constraints $t_{1,2} = t_{2,3}$, $\epsilon_2 = 0$, and $\epsilon_3 = -\epsilon_1$.

\subsection{Ground State Energy Convergence}
\begin{figure*}[t!] 
    \centering
    \begin{subfigure}[b]{0.32\textwidth} 
        \centering
        \includegraphics[width=\textwidth]{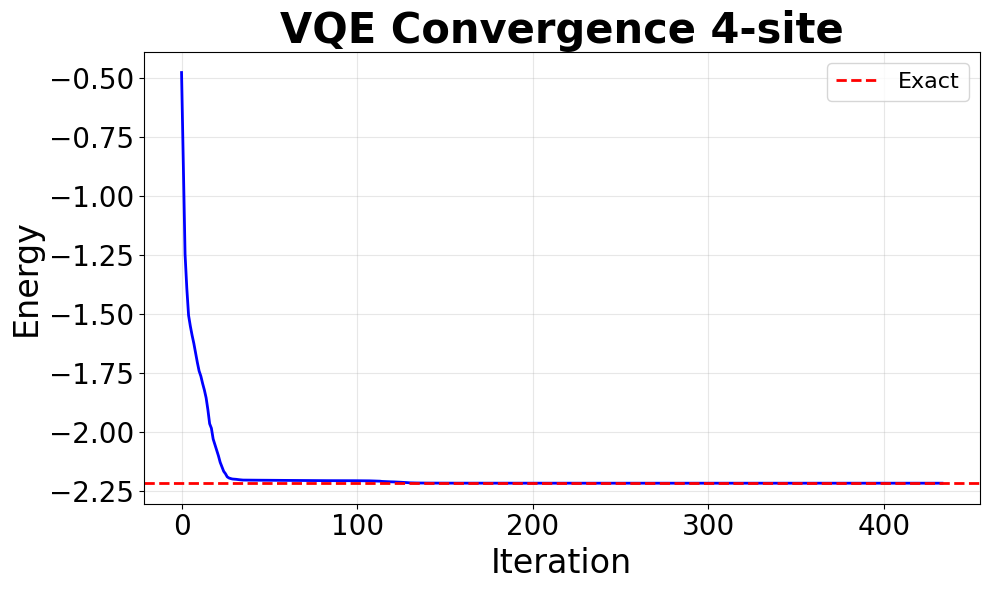}
        \caption{$U=4$}
        \label{fig:GS_U4}
    \end{subfigure}
    \hfill 
    \begin{subfigure}[b]{0.32\textwidth}
        \centering
        \includegraphics[width=\linewidth]{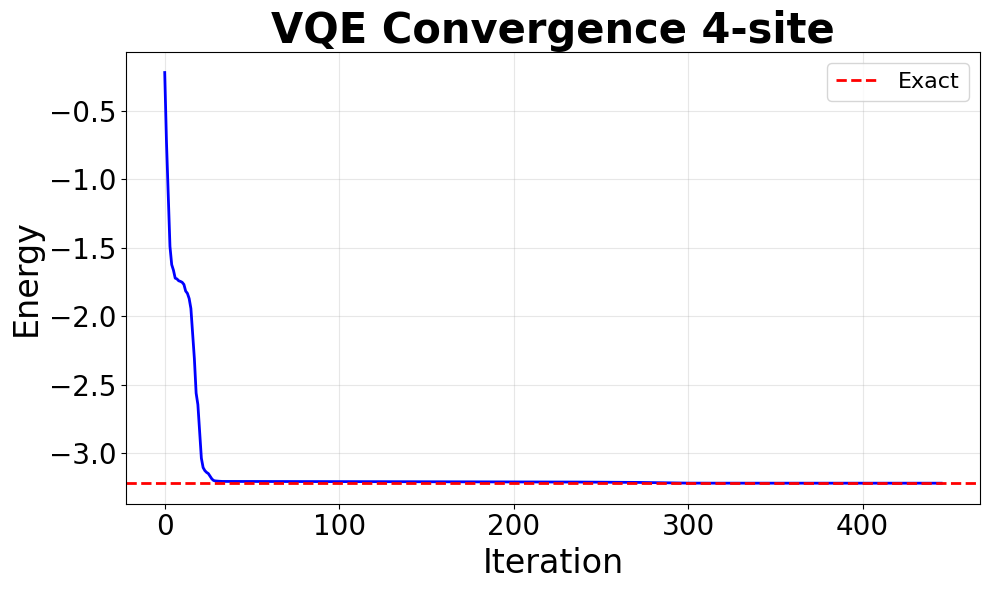}
        \caption{$U=6$}
        \label{fig:GS_U6}
    \end{subfigure}
    \begin{subfigure}[b]{0.32\textwidth}
        \centering
        \includegraphics[width=\linewidth]{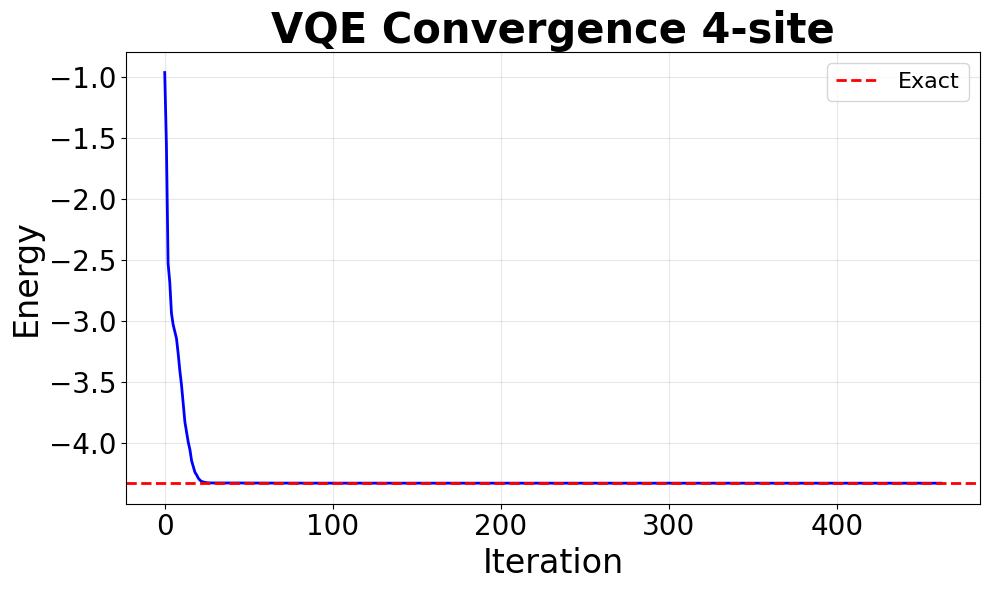}
        \caption{$U=8$}
        \label{fig:GS_U8}
    \end{subfigure}
    \label{fig:GSE}
    \hfill
    
    \caption{Convergence of the ground state energy as a function of the number of iterations for the VQE ansatz shown in Fig. \ref{fig:VQE_circuit}. Results are displayed for interaction strengths $U = 4,6, \text{and } 8$.}
    \label{fig:4site_VQE}
\end{figure*}

The variational parameters in the VQE circuit are optimized using the L-BFGS-B algorithm, a quasi-Newton method well-suited for bounded optimization problems \cite{Byrd_etal_1995}. Figure \ref{fig:4site_VQE} shows the convergence of the ground state energy as a function of iteration number for all three interaction strengths. We observe  convergence within approximately 100-150 iterations across all cases, with the optimization proceeding smoothly and avoiding significant plateaus or local minima traps. Table \ref{table:4site_error} summarizes the final optimized ground state energies compared with exact-diagonalization benchmarks. The VQE results achieve excellent agreement with the exact values. The relative error, defined as $ |E_{\text{exact}} -  E_{\text{VQE}}|/|E_{\text{exact}}|$ remains extremely small, reaching a maximum of only $0.006\%$ for the $U=4$ case. The slightly larger error at $U=4$ likely reflects the increased ground-state entanglement entropy characteristic of the metallic regime, which can be more challenging for parameterized circuits to capture \cite{Baul_etal_2025}. Nevertheless, even in this regime, the VQE approach maintains high accuracy, demonstrating the effectiveness of the symmetry-adapted ansatz. 

\begin{table}[b]
    \centering
    \begin{tabular}{c c c c}
        \toprule
        $U$ & Exact (ED) & VQE & Relative Error \\
        \midrule
        4 & -2.216536 & -2.216408 &  0.006\% \\
        6 & -3.219954& -3.219950&  0.0004\% \\
        8 & -4.330397 & -4.330377 & 0.0019\% \\
        \bottomrule
    \end{tabular}
    \caption{Comparison of ground state energies obtained from ED and VQE for different interaction strengths $U$, along with the corresponding relative errors.}
    \label{table:4site_error}
\end{table}

\subsection{Real-Time Retarded Green's Function}

\begin{figure*}[ht] 
    \centering
    \begin{subfigure}[b]{0.32\textwidth} 
        \centering
        \includegraphics[width=\textwidth]{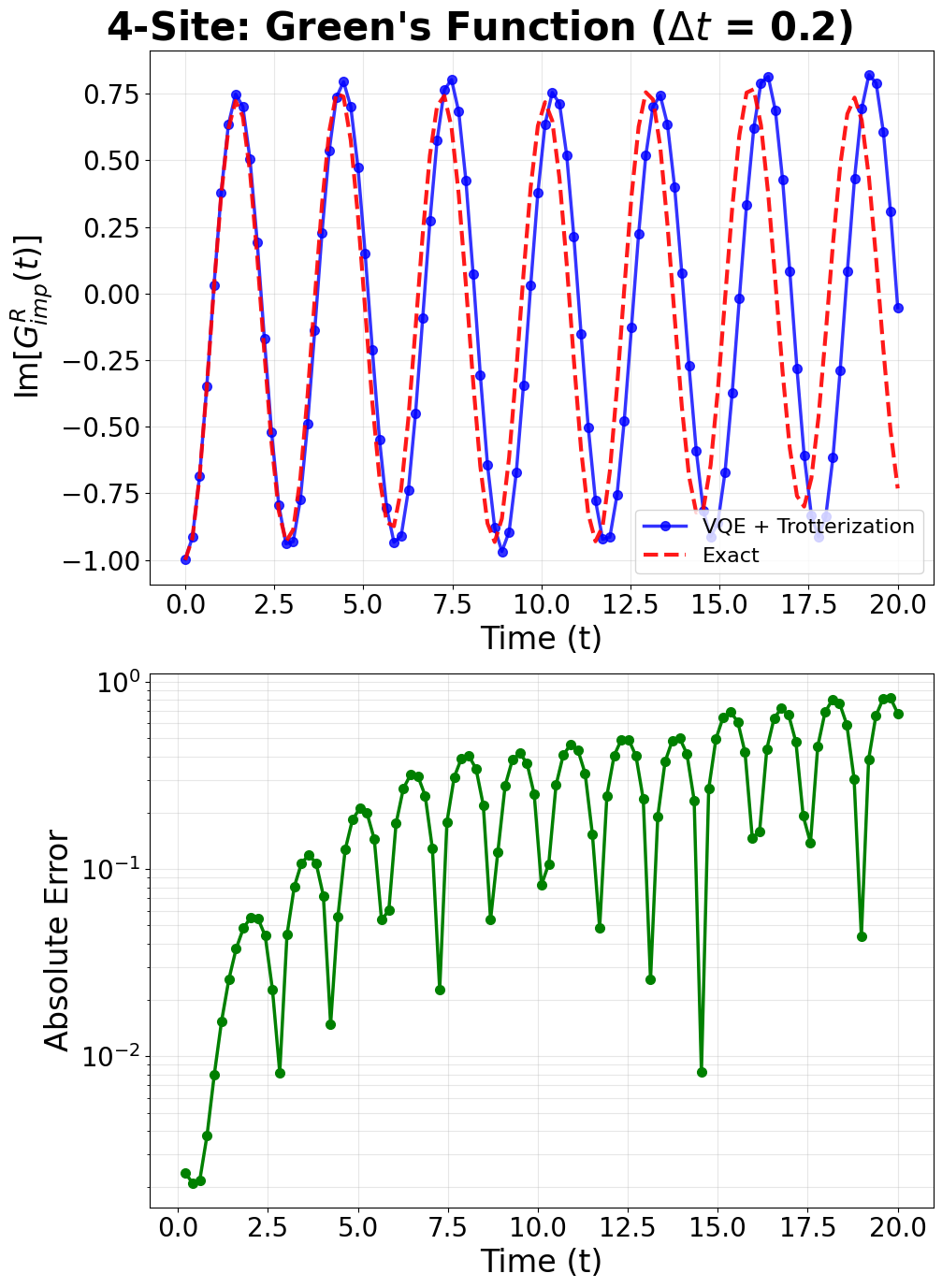}
        \caption{$U=4$}
        \label{fig:GF_dt0.2_U4}
    \end{subfigure}
    \hfill 
    \begin{subfigure}[b]{0.32\textwidth}
        \centering
        \includegraphics[width=\linewidth]{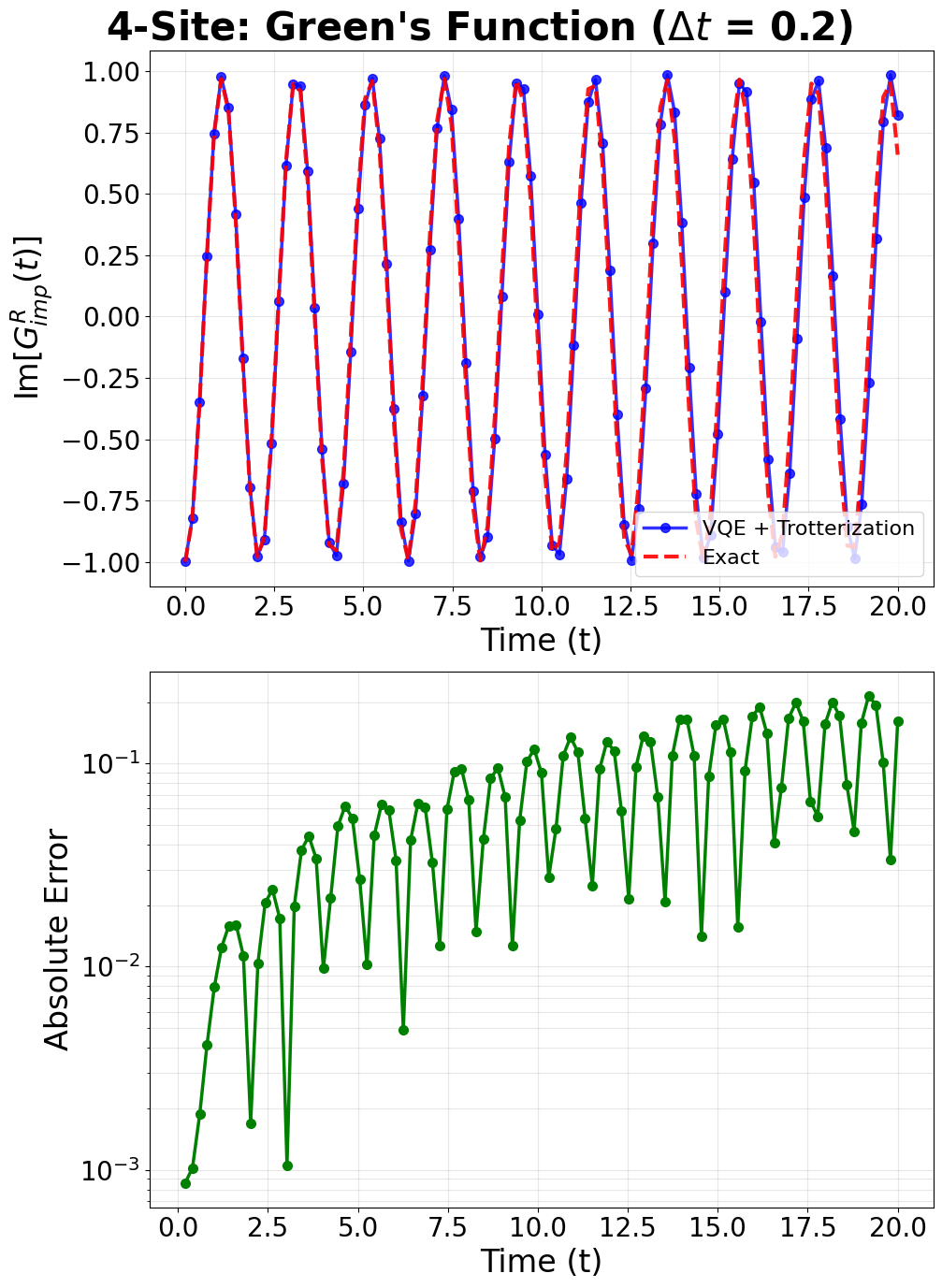}
        \caption{$U=6$}
        \label{fig:4b}
    \end{subfigure}
    \begin{subfigure}[b]{0.32\textwidth}
        \centering
        \includegraphics[width=\linewidth]{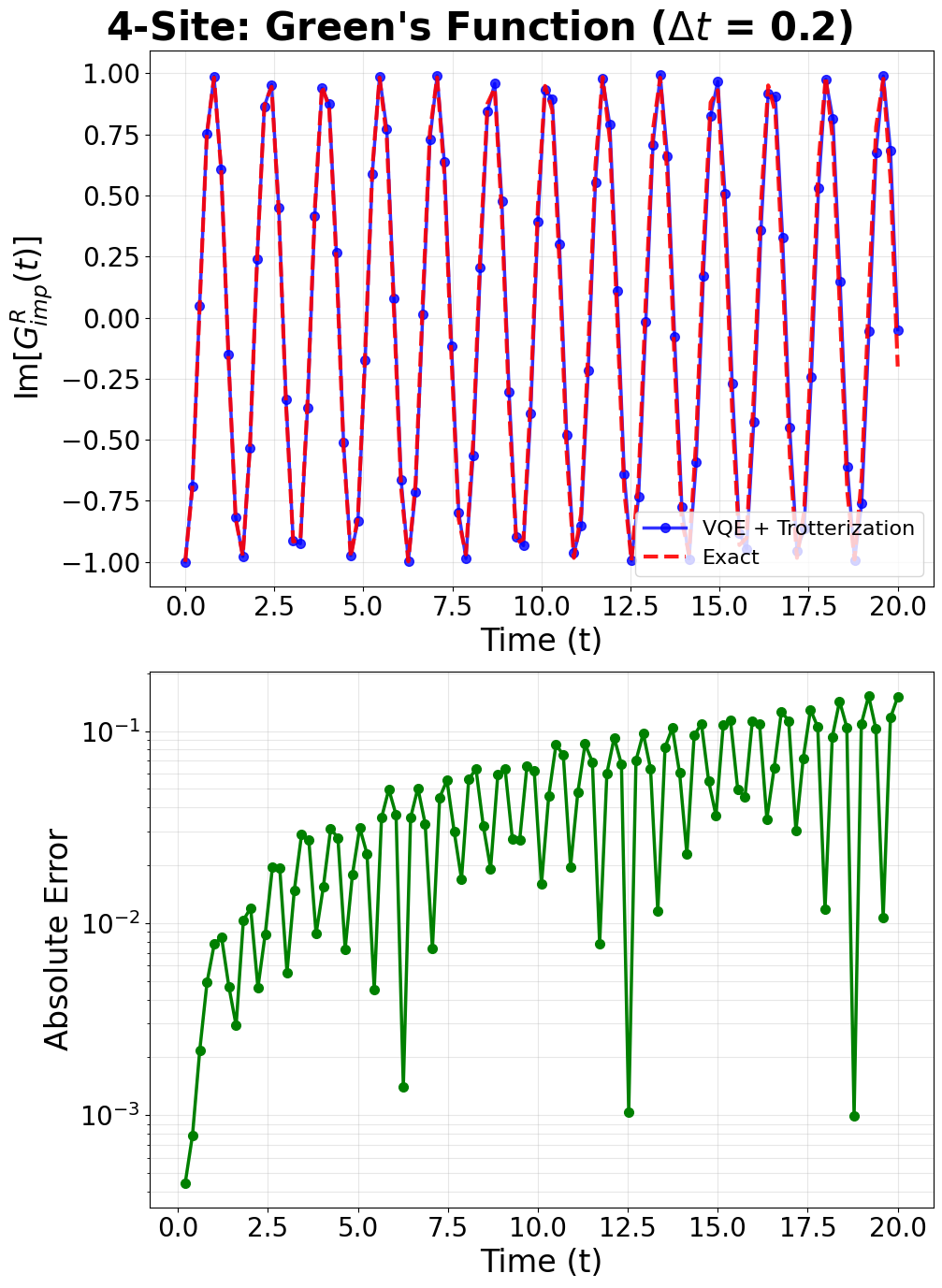}
        \caption{$U=8$}
        \label{fig:4c}
    \end{subfigure}
    \hfill
    
    \caption{Time evolution of the imaginary part of the retarded Green's function computed using the VQE-optimized ground state and Trotter propagation with a fixed time step $\Delta t=0.2$. The upper panels compare the VQE results with exact benchmarks for interaction strengths $U=4,6, \text{and } 8$. The lower panels display the corresponding absolute error of the VQE calculation relative to the exact solution.}
    \label{fig:4site_GF}
\end{figure*}

Having established the accuracy of the VQE ground state, we proceed to calculate the single-particle retarded Green's function. Using the VQE-optimized ground state as the initial condition, we compute the retarded Green's function in real time via Trotter time propagation with a fixed time step $\Delta t= 0.2$, following the formalism described in Eq. \ref{eq:GR}. At half-filling, particle-hole symmetry ensures that the real part of the Green's function vanishes identically; consequently, we focus exclusively on the imaginary part, $\text{Im}[G^R(t)]$, which encodes the spectral information of the system. Figure \ref{fig:4site_GF} displays the time evolution of the imaginary part of the Green's function (upper panels) alongside the absolute error relative to exact benchmarks (lower panels) for all three interaction strengths. The benchmark results were computed using the ground state from Exact Diagonalization (ED) evolved under the exact unitary operator. 

\begin{figure*}[t] 
    \centering
    \begin{subfigure}[b]{0.32\textwidth} 
        \centering
        \includegraphics[width=\textwidth]{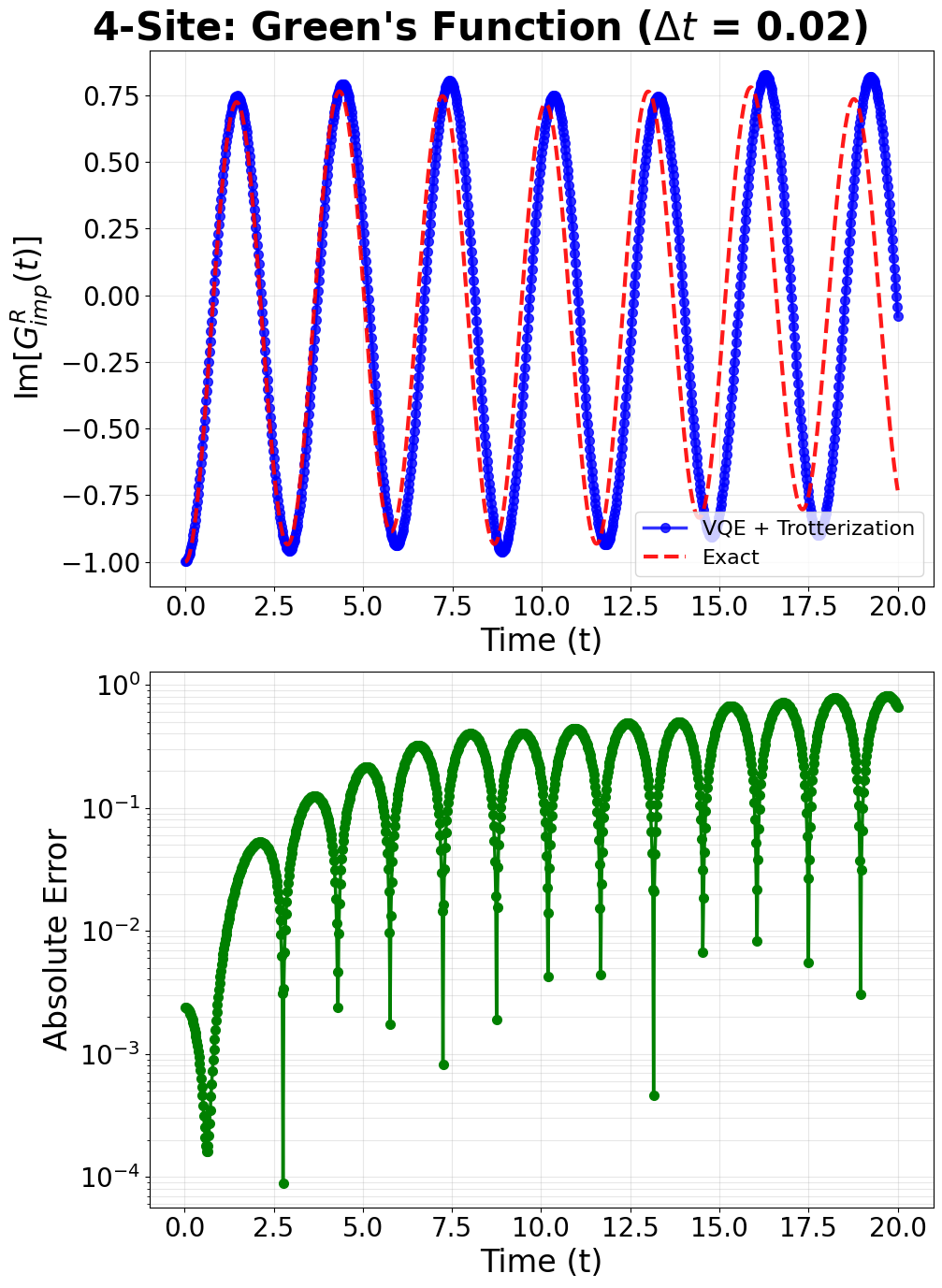}
        \caption{$U=4$}
        \label{fig:6a}
    \end{subfigure}
    \hfill 
    \begin{subfigure}[b]{0.32\textwidth}
        \centering
        \includegraphics[width=\linewidth]{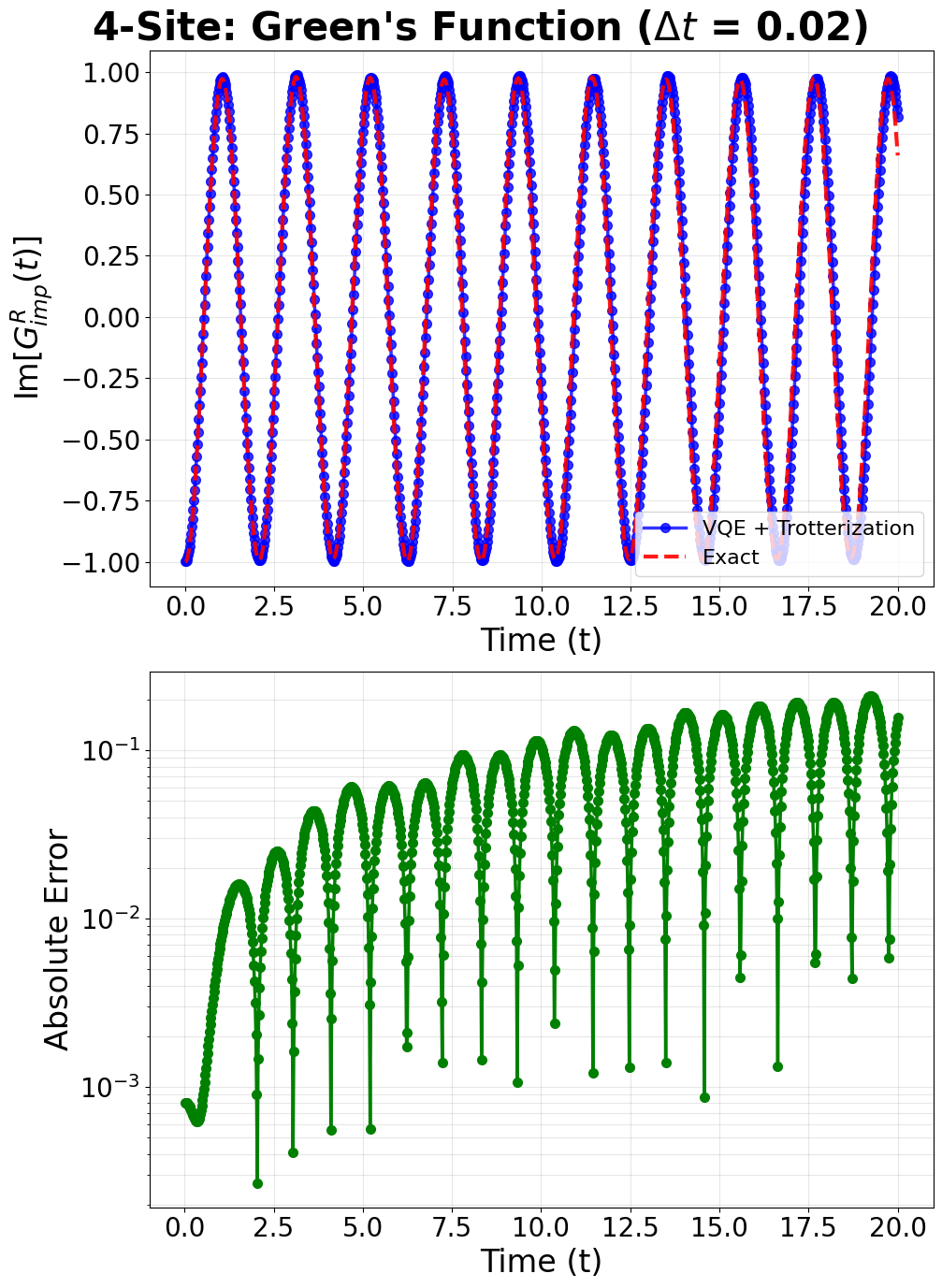}
        \caption{$U=6$}
        \label{fig:6b}
    \end{subfigure}
    \begin{subfigure}[b]{0.32\textwidth}
        \centering
        \includegraphics[width=\linewidth]{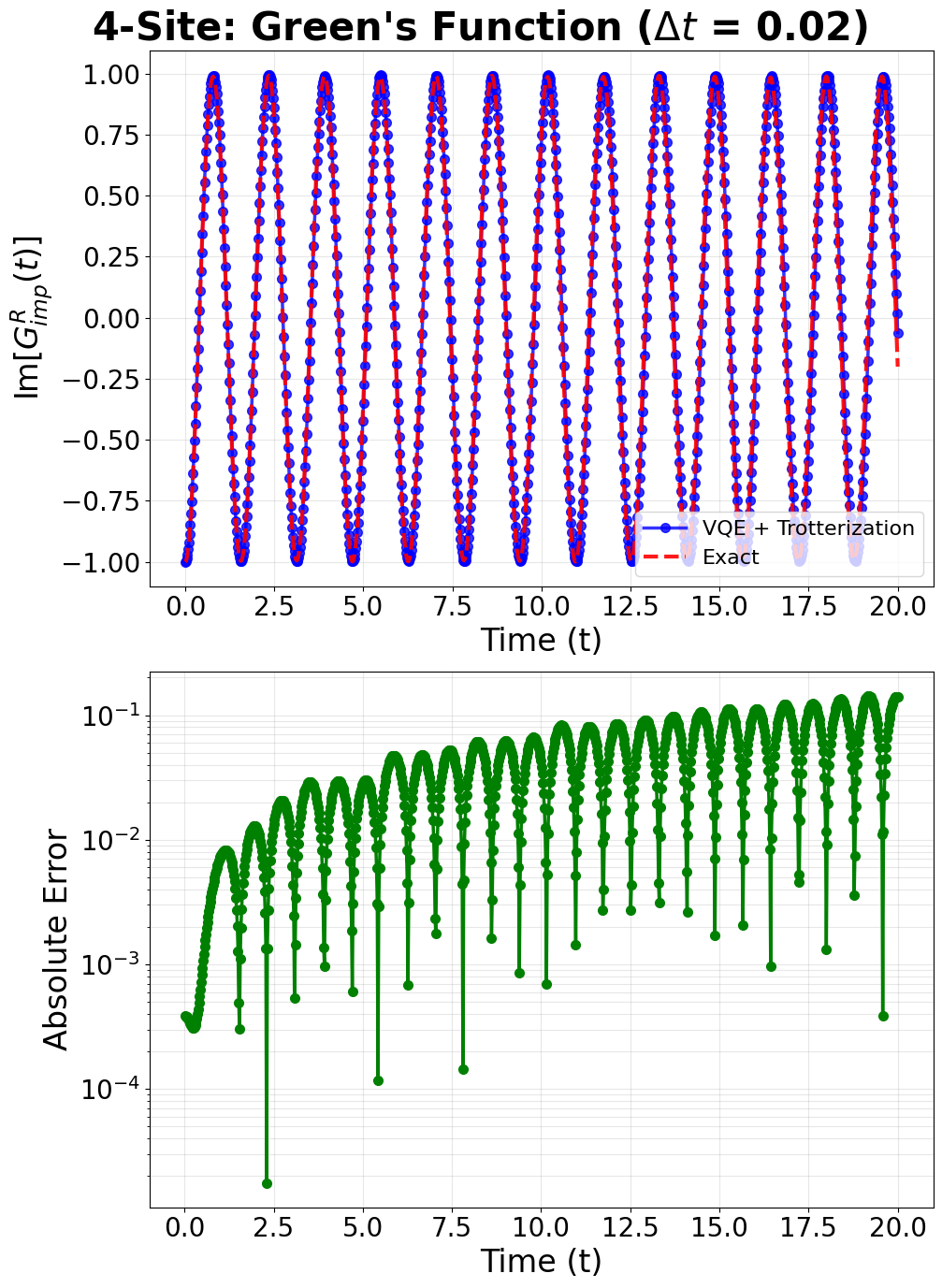}
        \caption{$U=8$}
        \label{fig:4site_GF_superfine}
    \end{subfigure}
    \hfill
    
    \caption{Time evolution of the imaginary part of the retarded Green's function computed using the VQE-optimized ground state and Trotter propagation with a refined time step $\Delta t=0.02$. The upper panels compare the VQE results with exact benchmarks for interaction strengths $U=4,6, \text{and } 8$, while the lower panels display the corresponding absolute error relative to the exact solution.}
    \label{fig:GF_superfine}
\end{figure*}

We observe good quantitative agreement between VQE and exact results across the entire simulated time domain ($0 \le t \le 20$). The VQE approach accurately captures the oscillatory structure of the Green's function, which encodes the system's characteristic excitation energies. The performance is particularly notable in the strongly correlated regimes $(U = 6, 8)$, demonstrating that the VQE ansatz successfully captures both the many-body correlations of the ground state and the transition matrix elements governing the dynamics. The error analysis in the lower panels of Fig. \ref{fig:4site_GF} reveals that deviations remain consistently low and oscillatory during short and intermediate time scales. At longer times, however, we observe a gradual accumulation of error, particularly in the $U=4$ regime. Notably, the error magnitude does not scale monotonically with interaction strength; instead, the weakly correlated metallic phase (U=4) exhibits larger long-time deviations compared to the more strongly correlated insulating cases ($U=6$ and $U=8$).

To disentangle the contributions of the approximate VQE ground state from the discretization error of the Trotter decomposition, we performed additional calculations with a significantly refined time step of $\Delta t=0.02$, a ten-fold reduction from our original value. As shown in Fig. \ref{fig:GF_superfine}, this refinement yields virtually identical error profiles to the coarser time step. This confirms that for the chosen step size of $\Delta t=0.02$, the dominant error source is the approximate nature of the VQE ground state rather than Trotter discretization. This finding has important practical implications: it suggests that reducing the time step further would incur higher circuit depths without much improvement in accuracy . Conversely, it validates that $\Delta t=0.2$ is sufficient to faithfully extract the dynamical information encoded in the current VQE wavefunction.


The enhanced error profile observed in the weak interaction regime ($U=4$) warrants further discussion. In this metallic phase, the spectral function is characterized by a quasiparticle peak at the Fermi level, characterized by a continuum of low-energy excitations. While the VQE optimization successfully minimizes the total ground state energy, it appears to struggle to capture correlations governing these low-lying modes. This highlights a significant challenge in variational approaches: minimizing the global energy expectation value does not strictly guarantee high fidelity for the entire excited state spectrum, particularly when energy differences are minute. The strongly correlated regimes ($U=6,8$) are characterized by large energy scales, where the system develops Hubbard bands separated by a gap of order U. These distinct, high-energy features are evidently more robustly captured by the VQE ansatz. One can also understand it through the lens of entanglement entropy. As detailed in Ref. \cite{Baul_etal_2025}, the entanglement in this impurity model decreases as a function of $U$; in the strong coupling limit, the formation of a rigid local moment on the impurity drives the system toward a factorizable product state, which is naturally easier for the parameterized circuit to represent. We note that calculations for weaker interactions ($U<4$) are not presented, as obtaining stable, self-consistent DMFT solutions for a three-bath discretization becomes numerically unstable in that regime. Based on the trend observed here, we anticipate that the accuracy of the VQE-computed Green's function would likely deteriorate further in the highly entangled, weak-coupling limit.

Despite these specific limitations, the overall agreement between the VQE and exact Green's functions across all correlation regimes indicates that the method is viable for practical DMFT applications, with the primary caveat being the accurate resolution of very low-lying excitations. Crucially, it is important to emphasize that these benchmarks employ realistic parameters derived from converged DMFT iterations, rather than artificially simplified test cases. This demonstrates the practical viability of the VQE approach for solving genuine impurity problems encountered in DMFT workflows.

\section{Conclusion}

In this work, we have demonstrated that the Variational Quantum Eigensolver (VQE), combined with real-time Trotter propagation, serves as an effective and accurate impurity solver for the Anderson Impurity Model (AIM) within the context of Dynamical Mean-Field Theory (DMFT). Our benchmarking against Exact Diagonalization (ED) for 4-site clusters (1 impurity coupled to 3 bath sites) across a range of interaction strengths ($U = 4, 6,$ and $8$) reveals that the VQE approach yields ground state energies with relative errors well below 0.01\%.

Several key findings emerge from this work. First, we confirm that constructing the variational ansatz with explicit symmetry preservation—specifically, total electron number conservation, total spin in the $z$-direction, and total spin symmetry—is crucial for achieving accurate ground state energies. This reinforces previous studies emphasizing the importance of symmetry-adapted ansätze for the quantum simulation of strongly correlated systems \cite{Jones_etal_2025,Gard_etal_2020}. Second, our results demonstrate that VQE-optimized ground states allow for accurate time evolution even when using a relatively coarse Trotter time step ($\Delta t=0.2$). The resulting Green's functions capture the essential dynamical information required for the DMFT self-consistency loop. The finding that a coarse time step is sufficient is practically significant, as it suggests that near-term hardware can extract dynamical properties without the prohibitive circuit depths required for extremely fine Trotterization.

The excellent agreement between VQE and exact results in the strongly correlated regime ($U = 6, 8$), indicates that this approach successfully captures the essential physics of the metal–insulator transition. Our work represents a step beyond the two-site DMFT approximation that has dominated early quantum computing implementations. While the two-site approximation can capture the onset of the transition via quasiparticle weights, it is fundamentally limited in its ability to reproduce detailed spectral features, as the Hilbert space contains only 4 accessible states once symmetries are enforced. By demonstrating that VQE works effectively for 4-site clusters, we open the door to recovering richer spectral information—such as distinct Hubbard bands—while remaining within the reach of near-term quantum hardware.

Even with an effective impurity solver, the success of the DMFT method depends on the convergence of the self-consistent loop. While standard DMFT implementations typically operate in Matsubara frequency or imaginary time, we have recently introduced a robust framework for performing the DMFT iteration directly in real time \cite{rangi2026}. This scheme has been successfully validated using exact diagonalization as the impurity solver, demonstrating stable convergence for the retarded Green's function. The critical next step is to integrate the VQE-based impurity solver presented here into this real-time self-consistency loop to verify that the approximate nature of the quantum solver does not destabilize the iteration.


Several promising directions for future research emerge from this study. First, extending benchmarks to larger clusters (e.g., 6 or more sites) is necessary to assess how solution accuracy scales with system size and to determine the bath discretization required for spectral convergence. Second, the full integration of this VQE-based solver with the real-time iteration scheme described in Ref. \cite{rangi2026} would enable end-to-end quantum-classical hybrid calculations, utilizing the quantum processor for the impurity problem while the classical computer manages the self-consistency. Third, investigating alternative ansatz architectures, such as hardware-efficient designs optimized for linear chain geometries or unitary coupled cluster (UCC) variants, could identify more resource-efficient strategies \cite{Peruzzo,Kandala_Nature_2017,Romero_QST_2018}. Finally, implementing this approach on actual quantum hardware will provide crucial insights into the impact of gate noise and decoherence on both state preparation and time evolution accuracy.

Beyond the single-band Hubbard model studied here, this methodology could be extended to multi-orbital systems and to cluster DMFT or extended DMFT formulations that incorporate spatial correlations \cite{Backes_etal_2023}. The capability to handle multiple orbitals would be particularly valuable for realistic material simulations, where $d$-band or $f$-band electrons often require multi-orbital treatments. Additionally, the real-time formulation employed here naturally suggests applications to non-equilibrium DMFT, enabling quantum computing studies of pump-probe experiments, transient dynamics, and other time-dependent phenomena in strongly correlated materials \cite{Freericks_2008,Aoki_etal_2014,Kreula_Clark_Jaksch_2016}. Furthermore, an encouraging recent finding is that the VQE method appears remarkably resilient to random disorder \cite{Alvertis_etal_2025}. This robustness suggests that the critical interplay between disorder and strong correlations could be effectively studied by combining the present impurity solver with the Coherent Potential Approximation (CPA) and related effective medium theories \cite{Dohner_etal_2022,Yan_Werner_2023,Rangi_Moreno_Tam_2024,Rangi2025Disorder,Abuelmaged_etal_2025,Soven_1967}.


In summary, we confirm that VQE combined with Trotter propagation provides a viable pathway for implementing DMFT impurity solvers on quantum computers. The method achieves good accuracy as compared to exact diagonalization for small clusters while remaining compatible with near-term quantum hardware constraints, particularly regarding the sufficiency of coarse time steps. These findings contribute to the growing body of evidence that quantum computing can play a meaningful role in solving strongly correlated electron problems, complementing classical methods and potentially extending the reach of DMFT to more complex materials than currently accessible. As quantum hardware continues to improve in qubit count, gate fidelity, and coherence times, the framework demonstrated here, combining symmetry-adapted VQE for ground state preparation with Trotter evolution for dynamical properties, represents a practical approach for studying strongly correlated quantum materials, although resolving low-energy excitations in correlated metals at weak to intermediate coupling remains a distinct challenge.

\begin{acknowledgments}
This work used high-performance computational resources provided by the Louisiana Optical Network Initiative and HPC@LSU computing.
\end{acknowledgments}



\bibliography{references}

\end{document}